\documentclass[%
 reprint,superscriptaddress,
amsmath,amssymb,
 aps,pre, 
]{revtex4-1}

\usepackage{siunitx}
\usepackage{subcaption}
\usepackage{xfrac}

\begin{document}

\title{Interplay of river and tidal forcings promotes loops in coastal channel networks}

\author{Adam Konkol}
\affiliation{Department of Physics and Astronomy, University of Pennsylvania, Philadelphia, PA, 19104}
\author{Jon Schwenk}
\affiliation{Earth and Environmental Sciences Division, Los Alamos National Laboratory}
\author{Eleni Katifori}
\affiliation{Department of Physics and Astronomy, University of Pennsylvania, Philadelphia, PA, 19104}
\author{John Burnham Shaw}
\affiliation{Department of Geosciences, University of Arkansas}
 
\date{\today}

\begin{abstract}
Global coastlines and their dense populations have an uncertain future due to increased flooding, storms, and human modification. The distributary channel networks of deltas and marshes that plumb these coastlines present diverse architectures, including well-studied dendritic topologies. However, the quasi-stable loops that are frequent in many coastal networks have not yet been explained. We present a model for self-organizing networks inspired by vascular biophysics to show that loops emerge when the relative forcings between rivers and tides are comparable, resulting in interplay between processes at short timescales relative to network evolution. Using field data and satellite imaging, we confirm this control on 21 natural networks. Our comparison provides the first evidence that hydrodynamic fluctuations promote loop formation in geophysical systems.
\end{abstract}

\pacs{Valid PACS appear here}

\maketitle

%How do coastal channel networks form, and what are their stable configurations? The answers to these questions are essential for predicting how these networks will evolve under growing human and climatic stresses. Loops exist in most coastal networks, yet physical controls on the initiation and stability of this fundamental network feature are unknown. We use a model of complex channel network evolution to show that loops form only when there is interplay between two distinct flow patterns. Field data show that river-tidal interplay can explain the presence or absence of loops on coastal networks that span Earth. This link between network structure and flow dynamics can indicate a network's equilibrium state or inform how it might react to changing conditions.}% Authors must submit a 120-word maximum statement about the significance of their research paper written at a level understandable to an undergraduate educated scientist outside their field of speciality. The primary goal of the significance statement is to explain the relevance of the work in broad context to a broad readership. The significance statement appears in the paper itself and is required for all research papers.

Coastal channel networks, which are the primary structural element of both river deltas and tidal estuaries, have sustained humans since the dawn of civilization \cite{bianchi_deltas_2016} and remain vital to growing populations today \cite{tessler_profiling_2015,edmonds_coastal_2020}. These networks are evolving - sometimes rapidly - due to altered water and sediment balances, human modification, and natural instabilities \cite{syvitski_sinking_2009,wilson_widespread_2017,dunn_projections_2019,nienhuis_global-scale_2020,schuerch_future_2018}. However, the complex topology of these networks \cite{tejedor2017} obscures the key drivers and emergent stable states \cite{passalacqua_delta_2017}. A clear understanding of channel network evolution is essential for a sustainable coastal future \cite{hoitink_resilience_2020}.

\begin{figure}[!ph]
    \centering
    \includegraphics[width=0.3\textwidth]{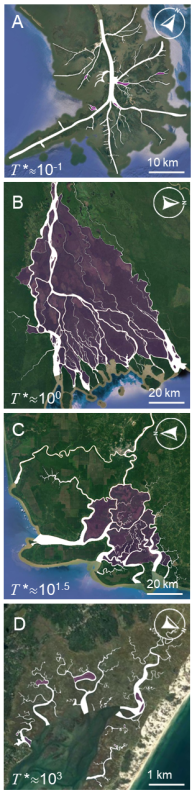}
    \caption{The (A) Mississippi, (B) Rajang, and (C) Orinoco Deltas and (D) Barnstable Marsh feature coastal networks spanning a wide range of $T^*$. Network masks are in white, and areas encompassed by loops are shaded purple. Satellite imaging from Google Earth.}
    % \phantomsubcaption \label{subfig:1a}
    % \phantomsubcaption \label{subfig:1b}
    % \phantomsubcaption \label{subfig:1c}
    % \phantomsubcaption \label{subfig:1d}
    \label{fig:1}
\end{figure}

Coastal channel networks distribute water and sediment between sources (e.g. a river or tidal inlet) and sinks (a bay, marsh, or coastal plain). Conceptual models of these networks feature channels that bifurcate recursively to form dendritic, tree-like structures that can distribute or collect water (as in Figure \ref{fig:1}A,\ref{fig:1}D). However, most coastal channel networks contain at least one loop, where alluvial channels formed by feedbacks between water and sediment transport branch and then rejoin around an island or shallow platform. We thus define loops as cycles in a channel network composed entirely of channels; not all islands are surrounded by loops as they may be partially or completely surrounded by unchannelized marsh, tidal flat, or ocean (Figure \ref{fig:1}). Transient loops may form from sandbar instabilities or avulsion processes that are not considered here \cite{parker_cause_1976,bolla_pittaluga_channel_2003,tejedor2017}, but many deltas contain looping networks (Figure \ref{fig:1}B,\ref{fig:1}C) that are stable over decades to centuries \cite{tejedor2015b,syvitski_anthropocene_2013,wilson_construction_2015,hoitink_tidal_2017}. 

The self-organized, tree-like structure found in some coastal channel networks may be rooted in tidal prism redistribution \cite{fagherazzi2008a} or mouth bar formation \cite{edmonds_quantitative_2011}. However, no morphodynamic explanations exists for the formation and persistence of loops within coastal channel networks. This knowledge gap is especially remarkable given the significant progress in the theory and modeling of dendritic channel network structures \cite{banavar2000,rinaldo_tidal_1999,fagherazzi_dynamics_2015,balister2018}.
%While the formation and topology of dendritic channel network structures are adequately explained and modeled \cite{banavar2000, fagherazzi_dynamics_2015, balister2018}, no morphodynamic explanation exists to explain the presence of persistent loops within coastal channel networks. Previous work has hinted at phenomenological suggestions that tidal prism re-distribution \cite{fagherazzi2008} or mouth bar formation \cite{edmonds_quantitative_2011} give rise to the self-organized, dendritic structure found in some coastal channel networks. 
%%%%%%%%%%%%%%%%%% Here point out grid-based modeling studies and how they miss loopy structure, or nod to tejedor's work on describing loopiness in real/modeled deltas. Keep the last sentence idea about structure and dynamics (although it seems like it could fit into the previous paragraph too)
%However, loop formation and persistence have not received the same scrutiny. 
Looping network topology has been statistically described \cite{passalacqua_geomorphic_2013, tejedor2015b}, but it has not been reproduced numerically or experimentally. Hence, our understanding of the feedback between structure and dynamics is limited in the presence of loops \cite{passalacqua_delta_2017}.  

%Optimal Channel Networks, theoretical networks optimized for minimal, uniform energy expenditure \cite{rodriguez1992a,rinaldo1992} adequately characterize many deltas and marsh networks \cite{rinaldo_tidal_1999,jimenez_scaling_2014}. However, Optimal Channel Networks as currently formulated have been proven to be loop free \cite{banavar2000,durand2007,balister2018}. This is because any loop will have one side shrink and one side grow because larger channels are capable of moving larger discharges more efficiently. Hence, our understanding of the interplay between structure and dynamics is limited in the presence of loops \cite{passalacqua_delta_2017}.  

Biological networks such as the vascular systems of animals or plants often feature loops that substantially increase resilience to damage or changes in flow \cite{katifori2010,corson2010,tero2010}. In these systems, it is believed that loops emerge and stabilize dynamically as a result of fluctuations in flow \cite{Hu2012,hu2013,ronellenfitsch2016}. In leaves, for example, loops in the vascular network appear to be related to spatiotemporal changes in auxin production during development \cite{Scarpella2006,ronellenfitsch2019}. Previous works have drawn comparisons between the structural features of biological and hydrological networks \cite{rodriguez1992b,pelletier2000,devauchelle2012,briggs2013}. Following clues from biological systems, we hypothesize that loops in coastal networks form as a result of fluctuations in channelized flow induced by river and tidal forcings. We test this with a numerical model of network formation wherein fluctuation intensity is systematically varied. We also empirically evaluate the relationship between fluctuations and loopiness across 21 natural coastal networks.

To quantify the interplay of river and tide forcings, we adopt the tide dominance ratio 
\begin{equation}
    T^*=Ah\omega/Q_r,
    \label{eq:tstar}
\end{equation} 
defined as the ratio of the characteristic tidal discharge to the mean annual flood discharge for tidal area $A$, average tidal range $h$, dominant tidal frequency $\omega$, and characteristic river flood discharge $Q_r$ \cite{nienhuis_future_2018}. The tide dominance ratio represents the strength of tidal inputs to the network relative to those of the river. We note that in the absence of $Q_r$ ($T^*\to \infty$), flood and ebb tidal currents are identical in magnitude and opposite in direction under this formulation, and this symmetry is broken for $Q_r>0$.
An inspection of river deltas reveals that loops emerge when the tide dominance ratio is near 1; considering $T^*$ alone makes it possible to continuously connect branched, loopless steady flow deltas to branched, loopless tidal marshes with a loopy intermediate state (Figure \ref{fig:1}).

\section*{Model Overview}
To investigate the controls on coastal network topology, we build a framework to model the evolution of an arbitrarily complex network of nodes and channels.
The flow $Q_e$ through an edge (or channel) $e$ is driven by the hydrostatic pressure drop $\Delta p_e$ via a linear relationship $Q_e= C_e  \Delta p_e / L_e$, with channel’s conductivity (inverse of bed-friction-derived flow resistivity) $C_e$ and channel length $L_e$. A river node supplies a constant discharge $Q_r$, and uniformly distributed tidal nodes throughout the network remove or add equal discharge sinusoidally in time to simulate a rising or falling tide. Nodes at the network’s coastal boundary ensure that water mass is conserved throughout the network by acting as sources or sinks according to the discharge and tidal flows. With knowledge of the network, time-dependent currents at the sources and sinks, and the conductances at each edge, we solve for the flows $Q_e$ throughout a tidal cycle using Kirchoff’s Laws (see Appendix). Tidal strength in each simulation is quantified as the ratio of the maximum net input current at the tidal nodes divided by current of the river node, directly related to $T^*$ defined for natural coastal networks.

Channels grow or shrink to adapt toward an equilibrium channel geometry and conductivity set by water and sediment discharge \cite{metivier_laboratory_2017,dunne_what_2020}. A simplified evolution equation \cite{tero2010,hu2013} captures the channel adaptation forces of deepening and widening via erosion (positive growth term) and shoaling and narrowing via sediment deposition (negative decay term) scaled by parameter $a$: 
\begin{equation}
\frac{dC_e}{dt}= a\langle Q_e^2 \rangle ^{\gamma} - \frac{1}{\tau}C_e, \label{eq:evo}
\end{equation} 
Taking tides to cycle much faster than the channel adaptation timescale $\tau$, we average the squared currents $\langle Q_e^2 \rangle$ over a tidal cycle. The empirically estimated exponent $\gamma=3/5$ ensures that the steady state $dC_e/dt=0$ conforms with the underlying scaling laws between channel properties (width $w_e$ and depth $d_e$) and average squared discharge estimated elsewhere \cite{myrick1963, bain2019}. This scaling law is derived to be $C_e = a\tau Q_e^{2\gamma}$ with conductivity $C_e\sim w_e d_e^2$ (\cite{hoitink_tidal_2017}; Appendix). Beginning with a random, planar network of nodes and channels (see SI), we solve for currents (see Appendix) and evolve the network using Equation \ref{eq:evo} until the network reaches equilibrium. Note that although the equilibrium, steady state networks produced by the adaptation equations used here are consistent with optimal channel networks \cite{rodriguez1992a,banavar2000,banavar2001,tejedor2017,balister2018}, here we do not explicitly require or demonstrate optimality of the final equilibrium network.

\section*{Results}
\begin{figure}[!ph]
    \centering
    \includegraphics[width=0.3\textwidth]{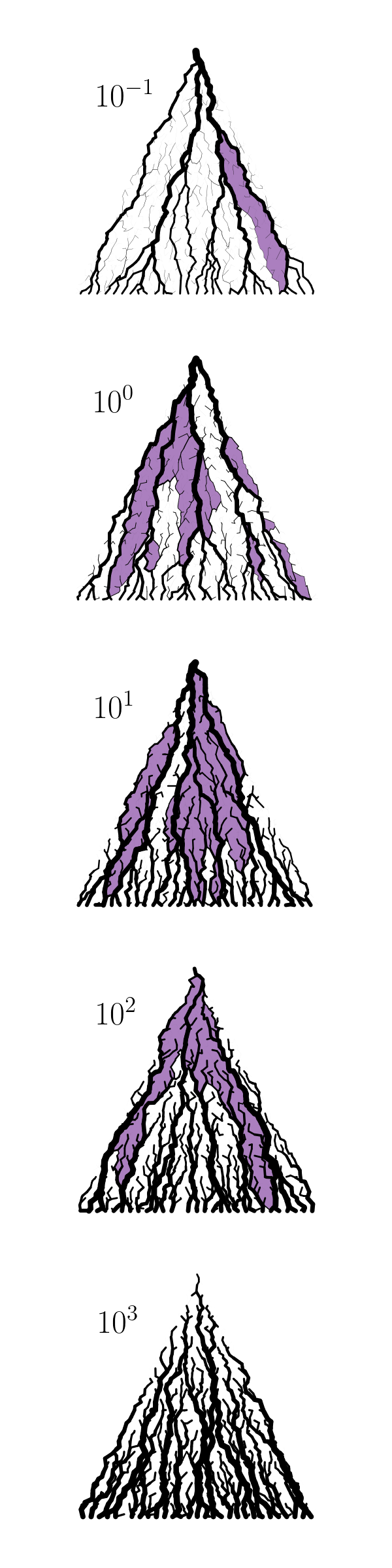}
    \caption{Simulated coastal channel networks with morphology spanning river dominated trees, intermediate loopy deltas, and loopless tidal marshes. The ratio of $T^*$ between maximum tidal inflow and river current is noted for each graph. Areas encompassed by loops are shaded purple.}
    \label{fig:2}
\end{figure}

%Not time dependence per se. In tidal end case, you have collective response of the system (dependent only on tidal phase). When you have both, you have frustration. No similarity to how the system responds. FRUSTRATION! ARRRG! Time dependence alone will not give you loops. 
%Model networks exhibit equilibrium topologies with loops at intermediate values of $T^*$ (Figure 2). In the presence of these loops we recognize the consequences of multiple temporally distinct forcings, while networks formed under either limit of river or tidal dominance confirm our analytical prediction of looplessness. At intermediate $T^*$, we interpret the interplay between these drivers of flow to arrest the channel narrowing feedback mechanism (sedimentation) that would otherwise cause at least one link in a loop to shrink indefinitely \cite{durand2007}. 
Simulated networks at steady state that are formed under either limit of river or tidal dominance lose any loops that existed in the initial condition. Our results agree with analytical prediction (SI Appendix) and established results for non-fluctuating channelized flow \cite{banavar2000,bohn2007}. Channels with increased conductivity along part of a loop carry water more efficiently, a feedback that causes the remainder of the loop to shrink indefinitely \cite{durand2007}. However, intermediate values of $T^*$ demand interplay between river and tidal forcings: $\langle Q_e^2 \rangle$ becomes a complex product of river current, bulk tidal flows, and the network structure. Networks with intermediate $T^*$ generally evolve toward stable states with loops within them (Figure \ref{fig:2}). 
We found this trend to be robust to changes in the density of tidal nodes, domain shape and tidal node distribution in our model (SI Appendix). 

%moved to discussion %Introducing more factors to increase flow variability through any given channel appears to further frustrate channel decay, as we tested by introducing spatial heterogeneity to tidal forcings to find loops persisting into high $T^*$ model networks. Under the conditions we test, these essential (unchanging over long times) fluctuations allow loops to exist at equilibrium (Figure 3). %Im not a fan of having a parenthetical remark after essential, or even the word essential (I wrote this myself for reference)
	
To test the model, we considered 21 coastal channel networks that span $T^*$, range in climate from tropical to polar, and vary in size from 0.2 - 6000 km$^2$. These include deltas in lakes with minimal tides (and thereby low $T^*$) and tidal networks with small river input (and thereby very high $T^*$). Binary channel masks of each network were created from overhead images or obtained from published sources (Supplementary Data File 1), and the channel networks including channel lengths and widths were extracted using RivGraph (\cite{schwenk2021}, SI Appendix). We sought to only compare loops at similar scales that are significant relative to the network. Hence, we removed all channels below a threshold (normalized to delta apex channel width) to account for unequal resolution of source images and network sizes. Both in model and natural networks, removing channels below a fixed threshold is akin to assuming that their flow is unchannelized, and our conclusions were consistent irrespective of threshold choice. 

\begin{figure}[!ph]
    \centering
    \includegraphics[width=0.49\textwidth]{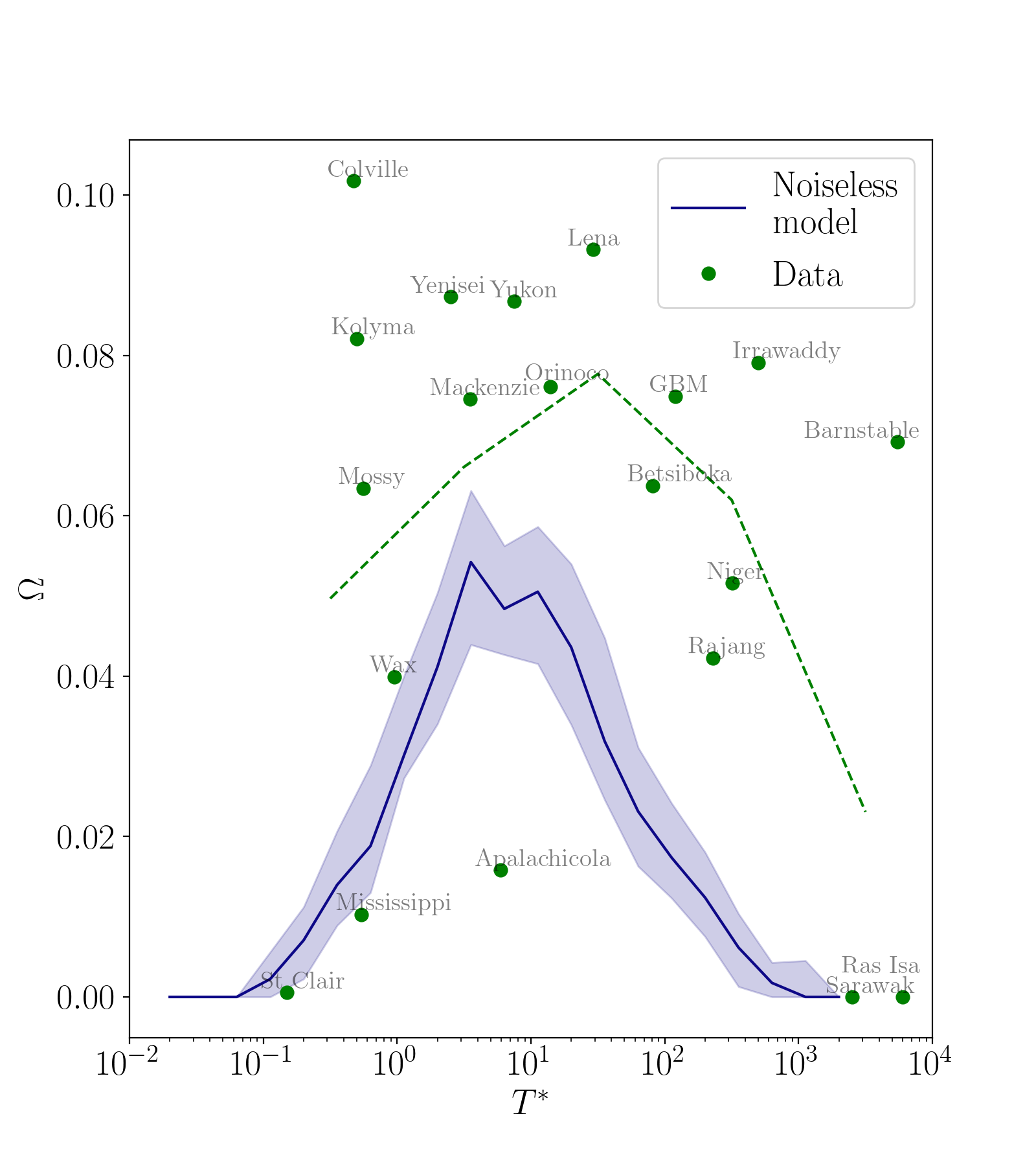}
    \includegraphics[width=0.49\textwidth]{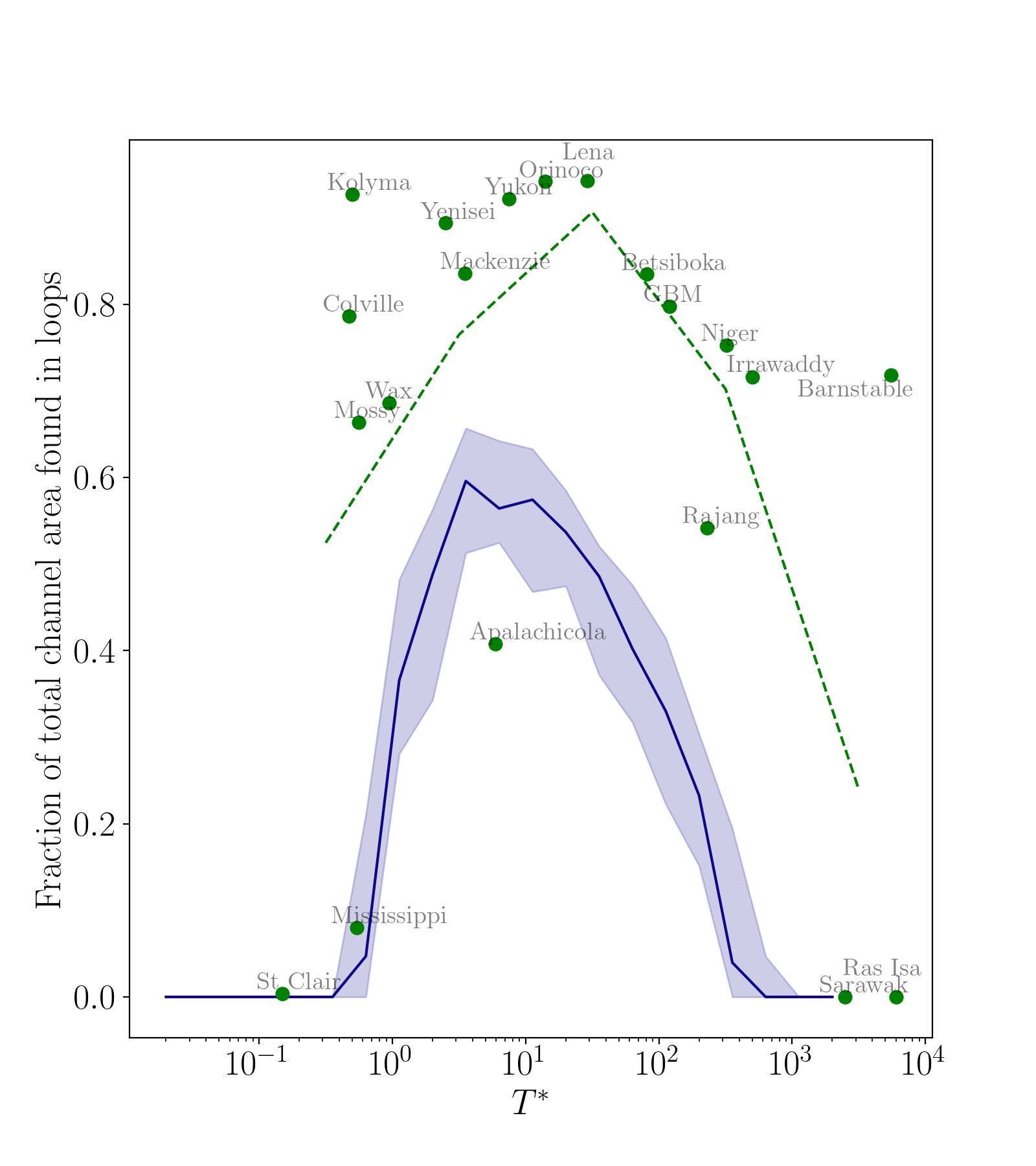}
    \caption{Measuring delta topology along an axis of $T^*$, the ratio of total tidal current to river current. Per-decade binned averages of data shown in green. (top, \ref{fig:3}A) The fraction of channel area found in loops on thresholded deltas and marshes (blue, median $\pm$ 25/75 percentile). (bottom, \ref{fig:3}B) Loopiness measured by $\Omega$, the minimum fraction of channel area that must be removed to make a topological tree. For initial random networks, $\Omega \approx 0.45$.}
    \label{fig:3}
    \phantomsubcaption\label{subfig:3a}
    \phantomsubcaption\label{subfig:3b}
\end{figure}

To compare modeled and real coastal channel networks, we identified independent dimensionless statistics that capture topological loopiness without knowledge of flows or dynamics. The fraction of total channel area comprising loops captures the expected effect of $T^*$ on network loopiness (Figure \ref{fig:3}A). Although intuitive, this measure identifies loops even if one of its channels is relatively insignificant. To complement this statistic, we also measure the minimum fraction of channel area that must be removed to make a tree ($\Omega$; Figure \ref{fig:3}B). This metric is robust against loops that include very narrow channels and quantifies how far complex networks are from dendritic end-member models.
%This measure, while intuitive, suffers from the binary, discontinuous nature of deciding which channels belong to a loop. Measuring the minimum fraction of channel area removed to make a tree ($\Omega$) gives a way to study the variation among loopy structures while removing dependnce on binary masks. 

A comparison of real and modeled coastal channel networks reveals a similar $T^*$ dependence of loops in both (Figure \ref{fig:3}). Natural loops are most prevalent for $T^*$ in the range $10^{0}-10^{2}$ with a peak near $10^1$ and least prevalent for $T^*<10^{-1}$ and $T^*>10^3$. To test the statistical significance of our findings, we randomized the $\Omega$ values of the 21 coastal networks. The randomized $\Omega$ plots had peaks as strong as the data no more than 3\% of the time when fit with piecewise linear, quadratic, or absolute value functions, indicating that the the observed trends in loopiness are not due to chance.

Broadly, the intermediate, loopy equilibrium morphology for tidal deltas presents a unique and previously undescribed prediction that is different from conventional scaling laws. Despite the direct relationship between tidal prism and delta area $A$, this trend was consistent when $A$ was controlled: the Mississippi, Kolyma, Mackenzie, Yukon, Orinoco, Betsiboka, and Rajang deltas vary in $A$ by a factor of only 4.8 (out of a total range of $\SI{3e4}{\square\kilo\meter}$) and still capture the trend in loopiness. In western Borneo, the loopy Rajang delta is found next to dendritic Sarawak channels, showing that networks found under similar climate and tidal forcing are still morphologically controlled by $T^*$.

\section*{Discussion}
Despite the model’s success in capturing the general dependence of coastal network topology on tides, modeled networks are overall less loopy than natural ones (Figure \ref{fig:3}), and there is variation between natural deltas with similar $T^*$ that the model does not yet capture. Our model represents perhaps the simplest representation of river-tide interplay within a network by linearizing flow, uniformly distributing flow across the tidal prism, and parameterizing water storage. The distortion of tidal waves through real networks produces additional spatiotemporal fluctuations observed in many coastal systems \cite{friedrichs_nonlinear_1992,bricheno_tidal_2016,hoitink_tidal_2016,hoitink_tidal_2017,mclachlan2020}. We considered such variations by adding random heterogeneity to the tidal discharges throughout the domain, which increased loopiness for all $T^*$ and became more significant with greater $T^*$. Even so, the maximum loopiness persisted in the intermediate range (SI Appendix 1.5). This shows that while river-tidal fluctuation is the dominant source of fluctuation, any process capable of inducing fluctuating hydrodynamics at short timescales compared to network evolution (e.g. vegetation dynamics, engineering practices) could potentially incite or suppress loops. We expect that further analysis into tidal heterogeneity, hydrodynamic nonlinearities and other anthropogenic or natural factors will capture more nuances in looping network structure and build upon the river-tidal control on loops in coastal networks shown here.

\section*{Conclusion}
Loops in coastal channel networks may stabilize under conditions where river-tidal interplay causes the time-averaged discharge to remain large enough to keep its channels open. The lack of this interplay (or unsteadiness caused by other heterogeneity) removes loops and produces simple trees. To our knowledge, this represents the first evidence in any natural system that topological loops can be stabilized by dynamically varying flow. Natural networks can be dominated by loops and generally contain at least one. We provide a framework for understanding how these complex networks organize that is simple and provides a foundation for analyzing network change under many drivers. This will provide new insight into the evolution of coastlines and its implications for coastal communities.

\begin{acknowledgments}
We thank Anastasia Piliouras for help with $T^*$ estimation for arctic deltas, and Maya Kedem for help mapping networks. \textbf{Funding:} Shaw was supported by a Department of Energy Grant (DESC0016163). Schwenk was supported by the Laboratory Directed Research and Development program of Los Alamos National Laboratory (project number 20170668PRD1). Katifori was supported by the NSF Career Award PHY-1554887 and the Simons Foundation through Award 568888. SI file is available upon request.  
\end{acknowledgments}

\appendix

\section*{Appendix}

\subsection*{Equilibrium networks}

Our key assumption is that channels are alluvial or self-formed, meaning that channels locally adjust to the characteristic water discharge they are subjected to. The assumption of local formation justifies our use of empirical power laws, which have been well established in coastal channel networks between characteristic discharge $Q$ and width $w$, depth $d$, and cross-sectionally averaged flow velocity $u$. We use relations $w\sim Q^\mu$ and $d\sim Q^\delta$ with $\mu = 0.5$ and $\delta=0.35$ (SI Appendix).

We take linear flow relations 

\begin{align*}
    Q&= wdu \\
    u &= \alpha \left( \frac{d}{L} \right) \Delta p
\end{align*}

\noindent for channel length $L$, end-to-end potential (water level) difference $\Delta p$, and proportionality constant $\alpha$. Substituting the latter equation into the former, we identify a conductivity (inverse resistance per unit length) $C=\alpha\;wd^2$ from the resulting Ohm’s law equation $Q=\alpha\;(wd^2/L)\Delta p$. Hence, channel conductivity scales as $C \sim Q^{\mu + 2\delta}$, such that we define $\gamma \equiv \mu/2 + \delta = 3/5$.

\subsection*{Solving currents} 

We use current (Neumann) boundary conditions to artificially simulate water levels rising and falling over a tidal cycle. If $\Delta$ is the network's oriented incidence matrix, Ohm's law reads 

\begin{align*}
    Q &= CL^{-1} \Delta p \\
\end{align*}
\noindent for diagonal matrices $C,L$ describing the conductivity and length of each edge, respectively. Using Kirchoff's current law with node source vector $S$ reading $\Delta^T Q = S$, we invert the graph Laplacian $\Delta^T CL^{-1}\Delta$ to solve for the pressures and, hence, the flows
\begin{align*}
    Q &= CL^{-1}\Delta(\Delta^T KL^{-1}\Delta)^\ddagger S.
\end{align*}

\noindent The double dagger indicates the Moore-Penrose pseudoinverse, as the graph Laplacian is singular. 

We used a node current source vector such that the river apex had constant current $\frac{1}{1+T^*/2}$ and the tidal nodes throughout the domain each had an equal share of the net tidal current $\frac{(T^*/2)\cos \omega t}{1+T^*/2}$. These currents were normalized so all model networks had maximum input current equal to $1$ at peak tidal discharge. Our use of $T^*$ here coincides with the definition in Equation \ref{eq:tstar}: the tidal volume $Ah$ changes in time with 

\begin{align*}
    \frac{d(Ah)}{dt} &= \frac{T^*/2}{1+T^*/2}\cos \omega t,
\end{align*}

\noindent which we integrate over flood tide ($0 < t < \pi/\omega$) to solve for $T^*$ as written in Equation \ref{eq:tstar}.

\subsection*{Data}
The dataset of coastal channel networks used for model comparison were selected to cover the full range of $T^*$ conditions. While river deltas naturally have $T^*$ between about $10^{-1}$ and $10^2$, we also include systems with higher $T^*$ are generally considered tidal channel networks due to their relatively inconsequential river inputs. Data collection involved (i) estimation of $T^*$ using data from field studies of each system, and (ii) careful network extraction. Estimates of flood discharge $Q_r$ and tidal range $h$ could generally be found in the published literature. Tidal area $A$ could be estimated from the mapped network and dominant tidal frequency was assumed to be the dominant diurnal frequency $\omega=(\sfrac{2\pi}{12.4})\si{\per\day}$.

The set of extracted channel networks analyzed here is the largest set of extracted networks assembled to date. See SI Apenndix Table 2 for details on the deltas and marshes we selected for study, and see SI Appendix Data File 1 for the data used. 

We use estimates of the tidal dominance ratio constrained with confidence to within a factor of four (plus or minus a factor of two) to reasonably place systems in logarithmic $T^*$ space (Equation \ref{eq:tstar}). The scientific literature and monitoring agencies generally constrain $h$, $\omega$, and $Q_r$. The area $A$ can be considered the full area of the delta within the channel network that can be nourished by water and sediment in its natural state \cite{edmonds_quantitative_2011}. While this includes the entire domain of channels and marshes of meso- and macro-tidal deltas and marshes, it generally does not include the island areas of micro-tidal arctic deltas (SI Appendix Table 2). The resulting values of characteristic tidal discharge $Ah\omega$ were generally larger than estimates of tidal discharge estimated from the morphology of channels \cite{nienhuis_future_2018,nienhuis_global-scale_2020}, sometimes by an order of magnitude. However, our choice is more suited to studying a suite of well-studied coastal systems, compared to the global analysis of hundreds or thousands of delta morphologies performed by other studies.

For extremely low $T^*$ systems, such as lacustrine deltas, and extremely high $T^*$ systems, found in tidal marshes, the tidal discharge or river discharge is often not measured and needs further estimation. While these estimates require larger uncertainty, they are done in order to place these end-member systems with logarithmic $T^*$ space. For the lacustrine deltas (Mossy and St. Clair), a tidal discharge proxy was set as was a characteristic lake level fluctuation (generally from seiches) over a 24 hour period recorded at a gauge. For the Ras Isa and Barnstable tidal marshes, river discharge was estimated as peak (90th percentile) rainfall within a network's watershed times the watershed area. We used the river discharge estimate in \cite{nienhuis_global-scale_2020} for the 5000 km$^2$ Sarawak tidal networks.

Binary, georeferenced river channel masks were created or obtained for each delta in our study (SI Appendix Table 2). In preparation for extracting the network topologies using the RivGraph Python package \cite{schwenk2021}, we first cropped the channel network masks to include only regions that were riverine or tidally influenced and eliminated portions of mask that were not connected to the delta channel network. Islands, which are represented as holes within the channel network mask, correspond to loops in the network and are therefore critically important to our analysis. Some islands in the masks were the result of noise in the base images or model used to generate the mask; these islands were typically on the order of a few pixels. We also sought to minimize islands from within-channel sediment bars. These are features arising from non-uniform flow acceleration and sediment transport \cite{seminara_sand_2001}, that are unlike the stable islands that are larger than one channel width that are studied here (Figure \ref{fig:2}). In order to eliminate, or fill, islands that do not contribute to the long-term network topology, we applied an automated thresholding scheme to each mask, detailed in the supplement.

The network of each channel network mask was then automatically extracted with RivGraph \cite{schwenk2021}, which returns the constituent links and nodes of the graph. RivGraph requires two additional inputs: a shapefile of inlet node locations and a shapefile of the shoreline. These were manually created for each network and are included in SI Appendix Data File 1. RivGraph was then used to prune each network to its shoreline and remove dangling links, and link widths and lengths were computed.

% Bibliography
\bibliography{RiversAndTheirLoopsArxiv}

\begin{thebibliography}{51}%
\makeatletter
\providecommand \@ifxundefined [1]{%
 \@ifx{#1\undefined}
}%
\providecommand \@ifnum [1]{%
 \ifnum #1\expandafter \@firstoftwo
 \else \expandafter \@secondoftwo
 \fi
}%
\providecommand \@ifx [1]{%
 \ifx #1\expandafter \@firstoftwo
 \else \expandafter \@secondoftwo
 \fi
}%
\providecommand \natexlab [1]{#1}%
\providecommand \enquote  [1]{``#1''}%
\providecommand \bibnamefont  [1]{#1}%
\providecommand \bibfnamefont [1]{#1}%
\providecommand \citenamefont [1]{#1}%
\providecommand \href@noop [0]{\@secondoftwo}%
\providecommand \href [0]{\begingroup \@sanitize@url \@href}%
\providecommand \@href[1]{\@@startlink{#1}\@@href}%
\providecommand \@@href[1]{\endgroup#1\@@endlink}%
\providecommand \@sanitize@url [0]{\catcode `\\12\catcode `\$12\catcode
  `\&12\catcode `\#12\catcode `\^12\catcode `\_12\catcode `\%12\relax}%
\providecommand \@@startlink[1]{}%
\providecommand \@@endlink[0]{}%
\providecommand \url  [0]{\begingroup\@sanitize@url \@url }%
\providecommand \@url [1]{\endgroup\@href {#1}{\urlprefix }}%
\providecommand \urlprefix  [0]{URL }%
\providecommand \Eprint [0]{\href }%
\providecommand \doibase [0]{http://dx.doi.org/}%
\providecommand \selectlanguage [0]{\@gobble}%
\providecommand \bibinfo  [0]{\@secondoftwo}%
\providecommand \bibfield  [0]{\@secondoftwo}%
\providecommand \translation [1]{[#1]}%
\providecommand \BibitemOpen [0]{}%
\providecommand \bibitemStop [0]{}%
\providecommand \bibitemNoStop [0]{.\EOS\space}%
\providecommand \EOS [0]{\spacefactor3000\relax}%
\providecommand \BibitemShut  [1]{\csname bibitem#1\endcsname}%
\let\auto@bib@innerbib\@empty
%</preamble>
\bibitem [{\citenamefont {Bianchi}(2016)}]{bianchi_deltas_2016}%
  \BibitemOpen
  \bibfield  {author} {\bibinfo {author} {\bibfnamefont {T.~S.}\ \bibnamefont
  {Bianchi}},\ }\href@noop {} {\emph {\bibinfo {title} {Deltas and {Humans}:
  {A} {Long} {Relationship} now {Threatened} by Global Change}}},\ \bibinfo
  {edition} {illustrated edition}\ ed.\ (\bibinfo  {publisher} {Oxford
  University Press},\ \bibinfo {address} {New York, NY},\ \bibinfo {year}
  {2016})\BibitemShut {NoStop}%
\bibitem [{\citenamefont {Tessler}\ \emph {et~al.}(2015)\citenamefont
  {Tessler}, \citenamefont {Vörösmarty}, \citenamefont {Grossberg},
  \citenamefont {Gladkova}, \citenamefont {Aizenman}, \citenamefont
  {Syvitski},\ and\ \citenamefont
  {Foufoula-Georgiou}}]{tessler_profiling_2015}%
  \BibitemOpen
  \bibfield  {author} {\bibinfo {author} {\bibfnamefont {Z.~D.}\ \bibnamefont
  {Tessler}}, \bibinfo {author} {\bibfnamefont {C.~J.}\ \bibnamefont
  {Vörösmarty}}, \bibinfo {author} {\bibfnamefont {M.}~\bibnamefont
  {Grossberg}}, \bibinfo {author} {\bibfnamefont {I.}~\bibnamefont {Gladkova}},
  \bibinfo {author} {\bibfnamefont {H.}~\bibnamefont {Aizenman}}, \bibinfo
  {author} {\bibfnamefont {J.~P.~M.}\ \bibnamefont {Syvitski}}, \ and\ \bibinfo
  {author} {\bibfnamefont {E.}~\bibnamefont {Foufoula-Georgiou}},\ }\href
  {\doibase 10.1126/science.aab3574} {\bibfield  {journal} {\bibinfo  {journal}
  {Science}\ }\textbf {\bibinfo {volume} {349}},\ \bibinfo {pages} {638}
  (\bibinfo {year} {2015})}\BibitemShut {NoStop}%
\bibitem [{\citenamefont {Edmonds}\ \emph {et~al.}(2020)\citenamefont
  {Edmonds}, \citenamefont {Caldwell}, \citenamefont {Brondizio},\ and\
  \citenamefont {Siani}}]{edmonds_coastal_2020}%
  \BibitemOpen
  \bibfield  {author} {\bibinfo {author} {\bibfnamefont {D.~A.}\ \bibnamefont
  {Edmonds}}, \bibinfo {author} {\bibfnamefont {R.~L.}\ \bibnamefont
  {Caldwell}}, \bibinfo {author} {\bibfnamefont {E.~S.}\ \bibnamefont
  {Brondizio}}, \ and\ \bibinfo {author} {\bibfnamefont {S.~M.~O.}\
  \bibnamefont {Siani}},\ }\href {\doibase 10.1038/s41467-020-18531-4}
  {\bibfield  {journal} {\bibinfo  {journal} {Nature Communications}\ }\textbf
  {\bibinfo {volume} {11}},\ \bibinfo {pages} {4741} (\bibinfo {year}
  {2020})}\BibitemShut {NoStop}%
\bibitem [{\citenamefont {Syvitski}\ \emph {et~al.}(2009)\citenamefont
  {Syvitski}, \citenamefont {Kettner}, \citenamefont {Overeem}, \citenamefont
  {Hutton}, \citenamefont {Hannon}, \citenamefont {Brakenridge}, \citenamefont
  {Day}, \citenamefont {Vörösmarty}, \citenamefont {Saito}, \citenamefont
  {Giosan},\ and\ \citenamefont {{others}}}]{syvitski_sinking_2009}%
  \BibitemOpen
  \bibfield  {author} {\bibinfo {author} {\bibfnamefont {J.~P.~M.}\
  \bibnamefont {Syvitski}}, \bibinfo {author} {\bibfnamefont {A.~J.}\
  \bibnamefont {Kettner}}, \bibinfo {author} {\bibfnamefont {I.}~\bibnamefont
  {Overeem}}, \bibinfo {author} {\bibfnamefont {E.~W.~H.}\ \bibnamefont
  {Hutton}}, \bibinfo {author} {\bibfnamefont {M.~T.}\ \bibnamefont {Hannon}},
  \bibinfo {author} {\bibfnamefont {G.~R.}\ \bibnamefont {Brakenridge}},
  \bibinfo {author} {\bibfnamefont {J.}~\bibnamefont {Day}}, \bibinfo {author}
  {\bibfnamefont {C.}~\bibnamefont {Vörösmarty}}, \bibinfo {author}
  {\bibfnamefont {Y.}~\bibnamefont {Saito}}, \bibinfo {author} {\bibfnamefont
  {L.}~\bibnamefont {Giosan}}, \ and\ \bibinfo {author} {\bibnamefont
  {{others}}},\ }\href {\doibase 10.1038/ngeo629} {\bibfield  {journal}
  {\bibinfo  {journal} {Nature Geoscience}\ }\textbf {\bibinfo {volume} {2}},\
  \bibinfo {pages} {681} (\bibinfo {year} {2009})}\BibitemShut {NoStop}%
\bibitem [{\citenamefont {Wilson}\ \emph {et~al.}(2017)\citenamefont {Wilson},
  \citenamefont {Goodbred}, \citenamefont {Small}, \citenamefont {Gilligan},
  \citenamefont {Sams}, \citenamefont {Mallick},\ and\ \citenamefont
  {Hale}}]{wilson_widespread_2017}%
  \BibitemOpen
  \bibfield  {author} {\bibinfo {author} {\bibfnamefont {C.}~\bibnamefont
  {Wilson}}, \bibinfo {author} {\bibfnamefont {S.}~\bibnamefont {Goodbred}},
  \bibinfo {author} {\bibfnamefont {C.}~\bibnamefont {Small}}, \bibinfo
  {author} {\bibfnamefont {J.}~\bibnamefont {Gilligan}}, \bibinfo {author}
  {\bibfnamefont {S.}~\bibnamefont {Sams}}, \bibinfo {author} {\bibfnamefont
  {B.}~\bibnamefont {Mallick}}, \ and\ \bibinfo {author} {\bibfnamefont
  {R.}~\bibnamefont {Hale}},\ }\href {\doibase 10.1525/elementa.263} {\bibfield
   {journal} {\bibinfo  {journal} {Elem Sci Anth}\ }\textbf {\bibinfo {volume}
  {5}} (\bibinfo {year} {2017}),\ 10.1525/elementa.263}\BibitemShut {NoStop}%
\bibitem [{\citenamefont {Dunn}\ \emph {et~al.}(2019)\citenamefont {Dunn},
  \citenamefont {Darby}, \citenamefont {Nicholls}, \citenamefont {Cohen},
  \citenamefont {Zarfl},\ and\ \citenamefont {Fekete}}]{dunn_projections_2019}%
  \BibitemOpen
  \bibfield  {author} {\bibinfo {author} {\bibfnamefont {F.~E.}\ \bibnamefont
  {Dunn}}, \bibinfo {author} {\bibfnamefont {S.~E.}\ \bibnamefont {Darby}},
  \bibinfo {author} {\bibfnamefont {R.~J.}\ \bibnamefont {Nicholls}}, \bibinfo
  {author} {\bibfnamefont {S.}~\bibnamefont {Cohen}}, \bibinfo {author}
  {\bibfnamefont {C.}~\bibnamefont {Zarfl}}, \ and\ \bibinfo {author}
  {\bibfnamefont {B.~M.}\ \bibnamefont {Fekete}},\ }\href {\doibase
  10.1088/1748-9326/ab304e} {\bibfield  {journal} {\bibinfo  {journal}
  {Environmental Research Letters}\ }\textbf {\bibinfo {volume} {14}},\
  \bibinfo {pages} {084034} (\bibinfo {year} {2019})}\BibitemShut {NoStop}%
\bibitem [{\citenamefont {Nienhuis}\ \emph {et~al.}()\citenamefont {Nienhuis},
  \citenamefont {Ashton}, \citenamefont {Edmonds}, \citenamefont {Hoitink},
  \citenamefont {Kettner}, \citenamefont {Rowland},\ and\ \citenamefont
  {Törnqvist}}]{nienhuis_global-scale_2020}%
  \BibitemOpen
  \bibfield  {author} {\bibinfo {author} {\bibfnamefont {J.~H.}\ \bibnamefont
  {Nienhuis}}, \bibinfo {author} {\bibfnamefont {A.~D.}\ \bibnamefont
  {Ashton}}, \bibinfo {author} {\bibfnamefont {D.~A.}\ \bibnamefont {Edmonds}},
  \bibinfo {author} {\bibfnamefont {A.~J.~F.}\ \bibnamefont {Hoitink}},
  \bibinfo {author} {\bibfnamefont {A.~J.}\ \bibnamefont {Kettner}}, \bibinfo
  {author} {\bibfnamefont {J.~C.}\ \bibnamefont {Rowland}}, \ and\ \bibinfo
  {author} {\bibfnamefont {T.~E.}\ \bibnamefont {Törnqvist}},\ }\href
  {https://www.nature.com/articles/s41586-019-1905-9} {\bibfield  {journal}
  {\bibinfo  {journal} {Nature}\ }\textbf {\bibinfo {volume}
  {577}}}\BibitemShut {NoStop}%
\bibitem [{\citenamefont {Schuerch}\ \emph {et~al.}(2018)\citenamefont
  {Schuerch}, \citenamefont {Spencer}, \citenamefont {Temmerman}, \citenamefont
  {Kirwan}, \citenamefont {Wolff}, \citenamefont {Lincke}, \citenamefont
  {McOwen}, \citenamefont {Pickering}, \citenamefont {Reef}, \citenamefont
  {Vafeidis}, \citenamefont {Hinkel}, \citenamefont {Nicholls},\ and\
  \citenamefont {Brown}}]{schuerch_future_2018}%
  \BibitemOpen
  \bibfield  {author} {\bibinfo {author} {\bibfnamefont {M.}~\bibnamefont
  {Schuerch}}, \bibinfo {author} {\bibfnamefont {T.}~\bibnamefont {Spencer}},
  \bibinfo {author} {\bibfnamefont {S.}~\bibnamefont {Temmerman}}, \bibinfo
  {author} {\bibfnamefont {M.~L.}\ \bibnamefont {Kirwan}}, \bibinfo {author}
  {\bibfnamefont {C.}~\bibnamefont {Wolff}}, \bibinfo {author} {\bibfnamefont
  {D.}~\bibnamefont {Lincke}}, \bibinfo {author} {\bibfnamefont {C.~J.}\
  \bibnamefont {McOwen}}, \bibinfo {author} {\bibfnamefont {M.~D.}\
  \bibnamefont {Pickering}}, \bibinfo {author} {\bibfnamefont {R.}~\bibnamefont
  {Reef}}, \bibinfo {author} {\bibfnamefont {A.~T.}\ \bibnamefont {Vafeidis}},
  \bibinfo {author} {\bibfnamefont {J.}~\bibnamefont {Hinkel}}, \bibinfo
  {author} {\bibfnamefont {R.~J.}\ \bibnamefont {Nicholls}}, \ and\ \bibinfo
  {author} {\bibfnamefont {S.}~\bibnamefont {Brown}},\ }\href {\doibase
  10.1038/s41586-018-0476-5} {\bibfield  {journal} {\bibinfo  {journal}
  {Nature}\ }\textbf {\bibinfo {volume} {561}},\ \bibinfo {pages} {231}
  (\bibinfo {year} {2018})}\BibitemShut {NoStop}%
\bibitem [{\citenamefont {Tejedor}\ \emph {et~al.}(2017)\citenamefont
  {Tejedor}, \citenamefont {Longjas}, \citenamefont {Edmonds}, \citenamefont
  {Zaliapin}, \citenamefont {Georgiou}, \citenamefont {Rinaldo},\ and\
  \citenamefont {Foufoula-Georgiou}}]{tejedor2017}%
  \BibitemOpen
  \bibfield  {author} {\bibinfo {author} {\bibfnamefont {A.}~\bibnamefont
  {Tejedor}}, \bibinfo {author} {\bibfnamefont {A.}~\bibnamefont {Longjas}},
  \bibinfo {author} {\bibfnamefont {D.~A.}\ \bibnamefont {Edmonds}}, \bibinfo
  {author} {\bibfnamefont {I.}~\bibnamefont {Zaliapin}}, \bibinfo {author}
  {\bibfnamefont {T.~T.}\ \bibnamefont {Georgiou}}, \bibinfo {author}
  {\bibfnamefont {A.}~\bibnamefont {Rinaldo}}, \ and\ \bibinfo {author}
  {\bibfnamefont {E.}~\bibnamefont {Foufoula-Georgiou}},\ }\href {\doibase
  10.1073/pnas.1708404114} {\bibfield  {journal} {\bibinfo  {journal} {of the
  National Academy of Sciences}\ }\textbf {\bibinfo {volume} {114}},\ \bibinfo
  {pages} {11651} (\bibinfo {year} {2017})}\BibitemShut {NoStop}%
\bibitem [{\citenamefont {Passalacqua}(2017)}]{passalacqua_delta_2017}%
  \BibitemOpen
  \bibfield  {author} {\bibinfo {author} {\bibfnamefont {P.}~\bibnamefont
  {Passalacqua}},\ }\href {\doibase 10.1016/j.geomorph.2016.04.001} {\bibfield
  {journal} {\bibinfo  {journal} {Geomorphology}\ }\bibinfo {series}
  {Connectivity in {Geomorphology} from {Binghamton} 2016},\ \textbf {\bibinfo
  {volume} {277}},\ \bibinfo {pages} {50} (\bibinfo {year} {2017})}\BibitemShut
  {NoStop}%
\bibitem [{\citenamefont {Hoitink}\ \emph {et~al.}(2020)\citenamefont
  {Hoitink}, \citenamefont {Nittrouer}, \citenamefont {Passalacqua},
  \citenamefont {Shaw}, \citenamefont {Langendoen}, \citenamefont {Huismans},\
  and\ \citenamefont {Maren}}]{hoitink_resilience_2020}%
  \BibitemOpen
  \bibfield  {author} {\bibinfo {author} {\bibfnamefont {A.~J.~F.}\
  \bibnamefont {Hoitink}}, \bibinfo {author} {\bibfnamefont {J.~A.}\
  \bibnamefont {Nittrouer}}, \bibinfo {author} {\bibfnamefont {P.}~\bibnamefont
  {Passalacqua}}, \bibinfo {author} {\bibfnamefont {J.~B.}\ \bibnamefont
  {Shaw}}, \bibinfo {author} {\bibfnamefont {E.~J.}\ \bibnamefont
  {Langendoen}}, \bibinfo {author} {\bibfnamefont {Y.}~\bibnamefont
  {Huismans}}, \ and\ \bibinfo {author} {\bibfnamefont {D.~S.~v.}\ \bibnamefont
  {Maren}},\ }\href {\doibase 10.1029/2019JF005201} {\bibfield  {journal}
  {\bibinfo  {journal} {Journal of Geophysical Research: Earth Surface}\ ,\
  \bibinfo {pages} {e2019JF005201}} (\bibinfo {year} {2020})}\BibitemShut
  {NoStop}%
\bibitem [{\citenamefont {Parker}(1976)}]{parker_cause_1976}%
  \BibitemOpen
  \bibfield  {author} {\bibinfo {author} {\bibfnamefont {G.}~\bibnamefont
  {Parker}},\ }\href {\doibase 10.1017/S0022112076000748} {\bibfield  {journal}
  {\bibinfo  {journal} {Journal of Fluid Mechanics}\ }\textbf {\bibinfo
  {volume} {76}},\ \bibinfo {pages} {457} (\bibinfo {year} {1976})}\BibitemShut
  {NoStop}%
\bibitem [{\citenamefont {Bolla~Pittaluga}\ \emph {et~al.}(2003)\citenamefont
  {Bolla~Pittaluga}, \citenamefont {Repetto},\ and\ \citenamefont
  {Tubino}}]{bolla_pittaluga_channel_2003}%
  \BibitemOpen
  \bibfield  {author} {\bibinfo {author} {\bibfnamefont {M.}~\bibnamefont
  {Bolla~Pittaluga}}, \bibinfo {author} {\bibfnamefont {R.}~\bibnamefont
  {Repetto}}, \ and\ \bibinfo {author} {\bibfnamefont {M.}~\bibnamefont
  {Tubino}},\ }\href@noop {} {\bibfield  {journal} {\bibinfo  {journal} {Water
  resources research}\ }\textbf {\bibinfo {volume} {39}},\ \bibinfo {pages}
  {1046} (\bibinfo {year} {2003})}\BibitemShut {NoStop}%
\bibitem [{\citenamefont {Tejedor}\ \emph {et~al.}(2015)\citenamefont
  {Tejedor}, \citenamefont {Longjas}, \citenamefont {Zaliapin},\ and\
  \citenamefont {Foufoula-Georgiou}}]{tejedor2015b}%
  \BibitemOpen
  \bibfield  {author} {\bibinfo {author} {\bibfnamefont {A.}~\bibnamefont
  {Tejedor}}, \bibinfo {author} {\bibfnamefont {A.}~\bibnamefont {Longjas}},
  \bibinfo {author} {\bibfnamefont {I.}~\bibnamefont {Zaliapin}}, \ and\
  \bibinfo {author} {\bibfnamefont {E.}~\bibnamefont {Foufoula-Georgiou}},\
  }\href {\doibase 10.1002/2014WR016604} {\bibfield  {journal} {\bibinfo
  {journal} {Water Resources Research}\ }\textbf {\bibinfo {volume} {51}},\
  \bibinfo {pages} {4019} (\bibinfo {year} {2015})}\BibitemShut {NoStop}%
\bibitem [{\citenamefont {Syvitski}\ \emph {et~al.}(2013)\citenamefont
  {Syvitski}, \citenamefont {Kettner}, \citenamefont {Overeem}, \citenamefont
  {Giosan}, \citenamefont {Brakenridge}, \citenamefont {Hannon},\ and\
  \citenamefont {Bilham}}]{syvitski_anthropocene_2013}%
  \BibitemOpen
  \bibfield  {author} {\bibinfo {author} {\bibfnamefont {J.~P.~M.}\
  \bibnamefont {Syvitski}}, \bibinfo {author} {\bibfnamefont {A.~J.}\
  \bibnamefont {Kettner}}, \bibinfo {author} {\bibfnamefont {I.}~\bibnamefont
  {Overeem}}, \bibinfo {author} {\bibfnamefont {L.}~\bibnamefont {Giosan}},
  \bibinfo {author} {\bibfnamefont {G.~R.}\ \bibnamefont {Brakenridge}},
  \bibinfo {author} {\bibfnamefont {M.}~\bibnamefont {Hannon}}, \ and\ \bibinfo
  {author} {\bibfnamefont {R.}~\bibnamefont {Bilham}},\ }\href {\doibase
  10.1016/j.ancene.2014.02.003} {\bibfield  {journal} {\bibinfo  {journal}
  {Anthropocene}\ }\textbf {\bibinfo {volume} {3}},\ \bibinfo {pages} {24}
  (\bibinfo {year} {2013})}\BibitemShut {NoStop}%
\bibitem [{\citenamefont {Wilson}\ and\ \citenamefont
  {Goodbred}(2015)}]{wilson_construction_2015}%
  \BibitemOpen
  \bibfield  {author} {\bibinfo {author} {\bibfnamefont {C.~A.}\ \bibnamefont
  {Wilson}}\ and\ \bibinfo {author} {\bibfnamefont {S.~L.}\ \bibnamefont
  {Goodbred}},\ }\href {\doibase 10.1146/annurev-marine-010213-135032}
  {\bibfield  {journal} {\bibinfo  {journal} {Annual Review of Marine Science}\
  }\textbf {\bibinfo {volume} {7}},\ \bibinfo {pages} {67} (\bibinfo {year}
  {2015})}\BibitemShut {NoStop}%
\bibitem [{\citenamefont {Hoitink}\ \emph {et~al.}(2017)\citenamefont
  {Hoitink}, \citenamefont {Wang}, \citenamefont {Vermeulen}, \citenamefont
  {Huismans},\ and\ \citenamefont {Kästner}}]{hoitink_tidal_2017}%
  \BibitemOpen
  \bibfield  {author} {\bibinfo {author} {\bibfnamefont {A.~J.~F.}\
  \bibnamefont {Hoitink}}, \bibinfo {author} {\bibfnamefont {Z.~B.}\
  \bibnamefont {Wang}}, \bibinfo {author} {\bibfnamefont {B.}~\bibnamefont
  {Vermeulen}}, \bibinfo {author} {\bibfnamefont {Y.}~\bibnamefont {Huismans}},
  \ and\ \bibinfo {author} {\bibfnamefont {K.}~\bibnamefont {Kästner}},\
  }\href {\doibase 10.1038/ngeo3000} {\bibfield  {journal} {\bibinfo  {journal}
  {Nature Geoscience}\ }\textbf {\bibinfo {volume} {10}},\ \bibinfo {pages}
  {637} (\bibinfo {year} {2017})}\BibitemShut {NoStop}%
\bibitem [{\citenamefont {Fagherazzi}(2008)}]{fagherazzi2008a}%
  \BibitemOpen
  \bibfield  {author} {\bibinfo {author} {\bibfnamefont {S.}~\bibnamefont
  {Fagherazzi}},\ }\href {\doibase 10.1073/pnas.0806668105} {\bibfield
  {journal} {\bibinfo  {journal} {of the National Academy of Sciences}\
  }\textbf {\bibinfo {volume} {105}},\ \bibinfo {pages} {18692} (\bibinfo
  {year} {2008})}\BibitemShut {NoStop}%
\bibitem [{\citenamefont {Edmonds}\ \emph {et~al.}(2011)\citenamefont
  {Edmonds}, \citenamefont {Paola}, \citenamefont {Hoyal},\ and\ \citenamefont
  {Sheets}}]{edmonds_quantitative_2011}%
  \BibitemOpen
  \bibfield  {author} {\bibinfo {author} {\bibfnamefont {D.}~\bibnamefont
  {Edmonds}}, \bibinfo {author} {\bibfnamefont {C.}~\bibnamefont {Paola}},
  \bibinfo {author} {\bibfnamefont {D.}~\bibnamefont {Hoyal}}, \ and\ \bibinfo
  {author} {\bibfnamefont {B.}~\bibnamefont {Sheets}},\ }\href {\doibase
  10.1029/2010JF001955} {\bibfield  {journal} {\bibinfo  {journal} {J. Geophys.
  Res}\ }\textbf {\bibinfo {volume} {116}},\ \bibinfo {pages} {F04022}
  (\bibinfo {year} {2011})}\BibitemShut {NoStop}%
\bibitem [{\citenamefont {Banavar}\ \emph {et~al.}(2000)\citenamefont
  {Banavar}, \citenamefont {Colaiori}, \citenamefont {Flammini}, \citenamefont
  {Maritan},\ and\ \citenamefont {Rinaldo}}]{banavar2000}%
  \BibitemOpen
  \bibfield  {author} {\bibinfo {author} {\bibfnamefont {J.~R.}\ \bibnamefont
  {Banavar}}, \bibinfo {author} {\bibfnamefont {F.}~\bibnamefont {Colaiori}},
  \bibinfo {author} {\bibfnamefont {A.}~\bibnamefont {Flammini}}, \bibinfo
  {author} {\bibfnamefont {A.}~\bibnamefont {Maritan}}, \ and\ \bibinfo
  {author} {\bibfnamefont {A.}~\bibnamefont {Rinaldo}},\ }\href {\doibase
  10.1103/PhysRevLett.84.4745} {\bibfield  {journal} {\bibinfo  {journal}
  {Phys. Rev. Lett.}\ }\textbf {\bibinfo {volume} {84}},\ \bibinfo {pages}
  {4745} (\bibinfo {year} {2000})}\BibitemShut {NoStop}%
\bibitem [{\citenamefont {Rinaldo}\ \emph {et~al.}(1999)\citenamefont
  {Rinaldo}, \citenamefont {Fagherazzi}, \citenamefont {Lanzoni}, \citenamefont
  {Marani},\ and\ \citenamefont {Dietrich}}]{rinaldo_tidal_1999}%
  \BibitemOpen
  \bibfield  {author} {\bibinfo {author} {\bibfnamefont {A.}~\bibnamefont
  {Rinaldo}}, \bibinfo {author} {\bibfnamefont {S.}~\bibnamefont {Fagherazzi}},
  \bibinfo {author} {\bibfnamefont {S.}~\bibnamefont {Lanzoni}}, \bibinfo
  {author} {\bibfnamefont {M.}~\bibnamefont {Marani}}, \ and\ \bibinfo {author}
  {\bibfnamefont {W.~E.}\ \bibnamefont {Dietrich}},\ }\href {\doibase
  199910.1029/1999WR900237} {\bibfield  {journal} {\bibinfo  {journal} {Water
  Resources Research}\ }\textbf {\bibinfo {volume} {35}},\ \bibinfo {pages} {P.
  3905} (\bibinfo {year} {1999})}\BibitemShut {NoStop}%
\bibitem [{\citenamefont {Fagherazzi}\ \emph {et~al.}(2015)\citenamefont
  {Fagherazzi}, \citenamefont {Edmonds}, \citenamefont {Nardin}, \citenamefont
  {Leonardi}, \citenamefont {Canestrelli}, \citenamefont {Falcini},
  \citenamefont {Jerolmack}, \citenamefont {Mariotti}, \citenamefont
  {Rowland},\ and\ \citenamefont {Slingerland}}]{fagherazzi_dynamics_2015}%
  \BibitemOpen
  \bibfield  {author} {\bibinfo {author} {\bibfnamefont {S.}~\bibnamefont
  {Fagherazzi}}, \bibinfo {author} {\bibfnamefont {D.}~\bibnamefont {Edmonds}},
  \bibinfo {author} {\bibfnamefont {W.}~\bibnamefont {Nardin}}, \bibinfo
  {author} {\bibfnamefont {N.}~\bibnamefont {Leonardi}}, \bibinfo {author}
  {\bibfnamefont {A.}~\bibnamefont {Canestrelli}}, \bibinfo {author}
  {\bibfnamefont {F.}~\bibnamefont {Falcini}}, \bibinfo {author} {\bibfnamefont
  {D.}~\bibnamefont {Jerolmack}}, \bibinfo {author} {\bibfnamefont
  {G.}~\bibnamefont {Mariotti}}, \bibinfo {author} {\bibfnamefont {J.~C.}\
  \bibnamefont {Rowland}}, \ and\ \bibinfo {author} {\bibfnamefont {R.~L.}\
  \bibnamefont {Slingerland}},\ }\href {\doibase 10.1002/2014RG000451}
  {\bibfield  {journal} {\bibinfo  {journal} {Reviews of Geophysics}\ ,\
  \bibinfo {pages} {2014RG000451}} (\bibinfo {year} {2015})}\BibitemShut
  {NoStop}%
\bibitem [{\citenamefont {Balister}\ \emph {et~al.}(2018)\citenamefont
  {Balister}, \citenamefont {Balogh}, \citenamefont {Bertuzzo}, \citenamefont
  {Bollob{\'a}s}, \citenamefont {Caldarelli}, \citenamefont {Maritan},
  \citenamefont {Mastrandrea}, \citenamefont {Morris},\ and\ \citenamefont
  {Rinaldo}}]{balister2018}%
  \BibitemOpen
  \bibfield  {author} {\bibinfo {author} {\bibfnamefont {P.}~\bibnamefont
  {Balister}}, \bibinfo {author} {\bibfnamefont {J.}~\bibnamefont {Balogh}},
  \bibinfo {author} {\bibfnamefont {E.}~\bibnamefont {Bertuzzo}}, \bibinfo
  {author} {\bibfnamefont {B.}~\bibnamefont {Bollob{\'a}s}}, \bibinfo {author}
  {\bibfnamefont {G.}~\bibnamefont {Caldarelli}}, \bibinfo {author}
  {\bibfnamefont {A.}~\bibnamefont {Maritan}}, \bibinfo {author} {\bibfnamefont
  {R.}~\bibnamefont {Mastrandrea}}, \bibinfo {author} {\bibfnamefont
  {R.}~\bibnamefont {Morris}}, \ and\ \bibinfo {author} {\bibfnamefont
  {A.}~\bibnamefont {Rinaldo}},\ }\href {\doibase 10.1073/pnas.1804484115}
  {\bibfield  {journal} {\bibinfo  {journal} {of the National Academy of
  Sciences of the United States of America}\ }\textbf {\bibinfo {volume}
  {115}},\ \bibinfo {pages} {6548} (\bibinfo {year} {2018})}\BibitemShut
  {NoStop}%
\bibitem [{\citenamefont {Passalacqua}\ \emph {et~al.}(2013)\citenamefont
  {Passalacqua}, \citenamefont {Lanzoni}, \citenamefont {Paola},\ and\
  \citenamefont {Rinaldo}}]{passalacqua_geomorphic_2013}%
  \BibitemOpen
  \bibfield  {author} {\bibinfo {author} {\bibfnamefont {P.}~\bibnamefont
  {Passalacqua}}, \bibinfo {author} {\bibfnamefont {S.}~\bibnamefont
  {Lanzoni}}, \bibinfo {author} {\bibfnamefont {C.}~\bibnamefont {Paola}}, \
  and\ \bibinfo {author} {\bibfnamefont {A.}~\bibnamefont {Rinaldo}},\ }\href
  {\doibase 10.1002/jgrf.20128} {\bibfield  {journal} {\bibinfo  {journal}
  {Journal of Geophysical Research: Earth Surface}\ }\textbf {\bibinfo {volume}
  {118}},\ \bibinfo {pages} {1838} (\bibinfo {year} {2013})}\BibitemShut
  {NoStop}%
\bibitem [{\citenamefont {Katifori}\ \emph {et~al.}(2010)\citenamefont
  {Katifori}, \citenamefont {Sz\"oll\ifmmode~\mbox{\H{o}}\else \H{o}\fi{}si},\
  and\ \citenamefont {Magnasco}}]{katifori2010}%
  \BibitemOpen
  \bibfield  {author} {\bibinfo {author} {\bibfnamefont {E.}~\bibnamefont
  {Katifori}}, \bibinfo {author} {\bibfnamefont {G.~J.}\ \bibnamefont
  {Sz\"oll\ifmmode~\mbox{\H{o}}\else \H{o}\fi{}si}}, \ and\ \bibinfo {author}
  {\bibfnamefont {M.~O.}\ \bibnamefont {Magnasco}},\ }\href {\doibase
  10.1103/PhysRevLett.104.048704} {\bibfield  {journal} {\bibinfo  {journal}
  {Phys. Rev. Lett.}\ }\textbf {\bibinfo {volume} {104}},\ \bibinfo {pages}
  {048704} (\bibinfo {year} {2010})}\BibitemShut {NoStop}%
\bibitem [{\citenamefont {Corson}(2010)}]{corson2010}%
  \BibitemOpen
  \bibfield  {author} {\bibinfo {author} {\bibfnamefont {F.}~\bibnamefont
  {Corson}},\ }\href {\doibase 10.1103/PhysRevLett.104.048703} {\bibfield
  {journal} {\bibinfo  {journal} {Phys. Rev. Lett.}\ }\textbf {\bibinfo
  {volume} {104}},\ \bibinfo {pages} {048703} (\bibinfo {year}
  {2010})}\BibitemShut {NoStop}%
\bibitem [{\citenamefont {Tero}\ \emph {et~al.}(2010)\citenamefont {Tero},
  \citenamefont {Takagi}, \citenamefont {Saigusa}, \citenamefont {Ito},
  \citenamefont {Bebber}, \citenamefont {Fricker}, \citenamefont {Yumiki},
  \citenamefont {Kobayashi},\ and\ \citenamefont {Nakagaki}}]{tero2010}%
  \BibitemOpen
  \bibfield  {author} {\bibinfo {author} {\bibfnamefont {A.}~\bibnamefont
  {Tero}}, \bibinfo {author} {\bibfnamefont {S.}~\bibnamefont {Takagi}},
  \bibinfo {author} {\bibfnamefont {T.}~\bibnamefont {Saigusa}}, \bibinfo
  {author} {\bibfnamefont {K.}~\bibnamefont {Ito}}, \bibinfo {author}
  {\bibfnamefont {D.~P.}\ \bibnamefont {Bebber}}, \bibinfo {author}
  {\bibfnamefont {M.~D.}\ \bibnamefont {Fricker}}, \bibinfo {author}
  {\bibfnamefont {K.}~\bibnamefont {Yumiki}}, \bibinfo {author} {\bibfnamefont
  {R.}~\bibnamefont {Kobayashi}}, \ and\ \bibinfo {author} {\bibfnamefont
  {T.}~\bibnamefont {Nakagaki}},\ }\href {\doibase 10.1126/science.1177894}
  {\bibfield  {journal} {\bibinfo  {journal} {Science}\ }\textbf {\bibinfo
  {volume} {327}},\ \bibinfo {pages} {439} (\bibinfo {year}
  {2010})}\BibitemShut {NoStop}%
\bibitem [{\citenamefont {Hu}\ \emph {et~al.}(2012)\citenamefont {Hu},
  \citenamefont {Cai},\ and\ \citenamefont {Rangan}}]{Hu2012}%
  \BibitemOpen
  \bibfield  {author} {\bibinfo {author} {\bibfnamefont {D.}~\bibnamefont
  {Hu}}, \bibinfo {author} {\bibfnamefont {D.}~\bibnamefont {Cai}}, \ and\
  \bibinfo {author} {\bibfnamefont {A.~V.}\ \bibnamefont {Rangan}},\ }\href
  {\doibase 10.1371/journal.pone.0045444} {\bibfield  {journal} {\bibinfo
  {journal} {PloS one}\ }\textbf {\bibinfo {volume} {7}},\ \bibinfo {pages}
  {e45444} (\bibinfo {year} {2012})}\BibitemShut {NoStop}%
\bibitem [{\citenamefont {Hu}\ and\ \citenamefont {Cai}(2013)}]{hu2013}%
  \BibitemOpen
  \bibfield  {author} {\bibinfo {author} {\bibfnamefont {D.}~\bibnamefont
  {Hu}}\ and\ \bibinfo {author} {\bibfnamefont {D.}~\bibnamefont {Cai}},\
  }\href {\doibase 10.1103/PhysRevLett.111.138701} {\bibfield  {journal}
  {\bibinfo  {journal} {Phys. Rev. Lett.}\ }\textbf {\bibinfo {volume} {111}},\
  \bibinfo {pages} {138701} (\bibinfo {year} {2013})}\BibitemShut {NoStop}%
\bibitem [{\citenamefont {Ronellenfitsch}\ and\ \citenamefont
  {Katifori}(2016)}]{ronellenfitsch2016}%
  \BibitemOpen
  \bibfield  {author} {\bibinfo {author} {\bibfnamefont {H.}~\bibnamefont
  {Ronellenfitsch}}\ and\ \bibinfo {author} {\bibfnamefont {E.}~\bibnamefont
  {Katifori}},\ }\href {\doibase 10.1103/PhysRevLett.117.138301} {\bibfield
  {journal} {\bibinfo  {journal} {Phys. Rev. Lett.}\ }\textbf {\bibinfo
  {volume} {117}},\ \bibinfo {pages} {138301} (\bibinfo {year}
  {2016})}\BibitemShut {NoStop}%
\bibitem [{\citenamefont {Scarpella}\ \emph {et~al.}(2006)\citenamefont
  {Scarpella}, \citenamefont {Marcos}, \citenamefont {Friml},\ and\
  \citenamefont {Berleth}}]{Scarpella2006}%
  \BibitemOpen
  \bibfield  {author} {\bibinfo {author} {\bibfnamefont {E.}~\bibnamefont
  {Scarpella}}, \bibinfo {author} {\bibfnamefont {D.}~\bibnamefont {Marcos}},
  \bibinfo {author} {\bibfnamefont {J.}~\bibnamefont {Friml}}, \ and\ \bibinfo
  {author} {\bibfnamefont {T.}~\bibnamefont {Berleth}},\ }\href {\doibase
  10.1101/gad.1402406} {\bibfield  {journal} {\bibinfo  {journal} {Genes and
  Development}\ }\textbf {\bibinfo {volume} {20}},\ \bibinfo {pages} {1015}
  (\bibinfo {year} {2006})}\BibitemShut {NoStop}%
\bibitem [{\citenamefont {Ronellenfitsch}\ and\ \citenamefont
  {Katifori}(2019)}]{ronellenfitsch2019}%
  \BibitemOpen
  \bibfield  {author} {\bibinfo {author} {\bibfnamefont {H.}~\bibnamefont
  {Ronellenfitsch}}\ and\ \bibinfo {author} {\bibfnamefont {E.}~\bibnamefont
  {Katifori}},\ }\href@noop {} {\bibfield  {journal} {\bibinfo  {journal}
  {Physical Review Letters}\ }\textbf {\bibinfo {volume} {123}} (\bibinfo
  {year} {2019})}\BibitemShut {NoStop}%
\bibitem [{\citenamefont {Rodríguez-Iturbe}\ \emph
  {et~al.}(1992{\natexlab{a}})\citenamefont {Rodríguez-Iturbe}, \citenamefont
  {Ijjász-Vásquez}, \citenamefont {Bras},\ and\ \citenamefont
  {Tarboton}}]{rodriguez1992b}%
  \BibitemOpen
  \bibfield  {author} {\bibinfo {author} {\bibfnamefont {I.}~\bibnamefont
  {Rodríguez-Iturbe}}, \bibinfo {author} {\bibfnamefont {E.~J.}\ \bibnamefont
  {Ijjász-Vásquez}}, \bibinfo {author} {\bibfnamefont {R.~L.}\ \bibnamefont
  {Bras}}, \ and\ \bibinfo {author} {\bibfnamefont {D.~G.}\ \bibnamefont
  {Tarboton}},\ }\href {\doibase 10.1029/91WR03033} {\bibfield  {journal}
  {\bibinfo  {journal} {Water Resources Research}\ }\textbf {\bibinfo {volume}
  {28}},\ \bibinfo {pages} {1089} (\bibinfo {year}
  {1992}{\natexlab{a}})}\BibitemShut {NoStop}%
\bibitem [{\citenamefont {Pelletier}\ and\ \citenamefont
  {Turcotte}(2000)}]{pelletier2000}%
  \BibitemOpen
  \bibfield  {author} {\bibinfo {author} {\bibfnamefont {J.~D.}\ \bibnamefont
  {Pelletier}}\ and\ \bibinfo {author} {\bibfnamefont {D.~L.}\ \bibnamefont
  {Turcotte}},\ }\href {\doibase 10.1098/rstb.2000.0566} {\bibfield  {journal}
  {\bibinfo  {journal} {Philosophical Transactions of the Royal Society of
  London. Series B: Biological Sciences}\ }\textbf {\bibinfo {volume} {355}},\
  \bibinfo {pages} {307} (\bibinfo {year} {2000})}\BibitemShut {NoStop}%
\bibitem [{\citenamefont {Devauchelle}\ \emph {et~al.}(2012)\citenamefont
  {Devauchelle}, \citenamefont {Petroff}, \citenamefont {Seybold},\ and\
  \citenamefont {Rothman}}]{devauchelle2012}%
  \BibitemOpen
  \bibfield  {author} {\bibinfo {author} {\bibfnamefont {O.}~\bibnamefont
  {Devauchelle}}, \bibinfo {author} {\bibfnamefont {A.~P.}\ \bibnamefont
  {Petroff}}, \bibinfo {author} {\bibfnamefont {H.~F.}\ \bibnamefont
  {Seybold}}, \ and\ \bibinfo {author} {\bibfnamefont {D.~H.}\ \bibnamefont
  {Rothman}},\ }\href {\doibase 10.1073/pnas.1215218109} {\bibfield  {journal}
  {\bibinfo  {journal} {of the National Academy of Sciences}\ }\textbf
  {\bibinfo {volume} {109}},\ \bibinfo {pages} {20832} (\bibinfo {year}
  {2012})}\BibitemShut {NoStop}%
\bibitem [{\citenamefont {Briggs}\ and\ \citenamefont
  {Krishnamoorthy}(2013)}]{briggs2013}%
  \BibitemOpen
  \bibfield  {author} {\bibinfo {author} {\bibfnamefont {L.~A.}\ \bibnamefont
  {Briggs}}\ and\ \bibinfo {author} {\bibfnamefont {M.}~\bibnamefont
  {Krishnamoorthy}},\ }\href {\doibase 10.1073/pnas.1313866110} {\bibfield
  {journal} {\bibinfo  {journal} {of the National Academy of Sciences}\
  }\textbf {\bibinfo {volume} {110}},\ \bibinfo {pages} {19295} (\bibinfo
  {year} {2013})}\BibitemShut {NoStop}%
\bibitem [{\citenamefont {Nienhuis}\ \emph {et~al.}(2018)\citenamefont
  {Nienhuis}, \citenamefont {Hoitink},\ and\ \citenamefont
  {Törnqvist}}]{nienhuis_future_2018}%
  \BibitemOpen
  \bibfield  {author} {\bibinfo {author} {\bibfnamefont {J.~H.}\ \bibnamefont
  {Nienhuis}}, \bibinfo {author} {\bibfnamefont {A.~J. F.~T.}\ \bibnamefont
  {Hoitink}}, \ and\ \bibinfo {author} {\bibfnamefont {T.~E.}\ \bibnamefont
  {Törnqvist}},\ }\href {\doibase 10.1029/2018GL077638} {\bibfield  {journal}
  {\bibinfo  {journal} {Geophysical Research Letters}\ }\textbf {\bibinfo
  {volume} {45}},\ \bibinfo {pages} {3499} (\bibinfo {year}
  {2018})}\BibitemShut {NoStop}%
\bibitem [{\citenamefont {Métivier}\ \emph {et~al.}(2017)\citenamefont
  {Métivier}, \citenamefont {Lajeunesse},\ and\ \citenamefont
  {Devauchelle}}]{metivier_laboratory_2017}%
  \BibitemOpen
  \bibfield  {author} {\bibinfo {author} {\bibfnamefont {F.}~\bibnamefont
  {Métivier}}, \bibinfo {author} {\bibfnamefont {E.}~\bibnamefont
  {Lajeunesse}}, \ and\ \bibinfo {author} {\bibfnamefont {O.}~\bibnamefont
  {Devauchelle}},\ }\href {\doibase 10.5194/esurf-5-187-2017} {\bibfield
  {journal} {\bibinfo  {journal} {Earth Surface Dynamics}\ }\textbf {\bibinfo
  {volume} {5}},\ \bibinfo {pages} {187} (\bibinfo {year} {2017})}\BibitemShut
  {NoStop}%
\bibitem [{\citenamefont {Dunne}\ and\ \citenamefont
  {Jerolmack}(2020)}]{dunne_what_2020}%
  \BibitemOpen
  \bibfield  {author} {\bibinfo {author} {\bibfnamefont {K.~B.~J.}\
  \bibnamefont {Dunne}}\ and\ \bibinfo {author} {\bibfnamefont {D.~J.}\
  \bibnamefont {Jerolmack}},\ }\href {\doibase 10.1126/sciadv.abc1505}
  {\bibfield  {journal} {\bibinfo  {journal} {Science Advances}\ }\textbf
  {\bibinfo {volume} {6}},\ \bibinfo {pages} {eabc1505} (\bibinfo {year}
  {2020})}\BibitemShut {NoStop}%
\bibitem [{\citenamefont {Myrick}\ and\ \citenamefont
  {Leopold}(1963)}]{myrick1963}%
  \BibitemOpen
  \bibfield  {author} {\bibinfo {author} {\bibfnamefont {R.~M.}\ \bibnamefont
  {Myrick}}\ and\ \bibinfo {author} {\bibfnamefont {L.~B.}\ \bibnamefont
  {Leopold}},\ }\href {\doibase 10.3133/pp422B} {\emph {\bibinfo {title}
  {Hydraulic geometry of a small tidal estuary}}},\ \bibinfo {type} {Tech.
  Rep.}\ (\bibinfo {address} {Washington, D.C.},\ \bibinfo {year} {1963})\
  \bibinfo {note} {report}\BibitemShut {NoStop}%
\bibitem [{\citenamefont {Bain}\ \emph {et~al.}(2019)\citenamefont {Bain},
  \citenamefont {Hale},\ and\ \citenamefont {Goodbred}}]{bain2019}%
  \BibitemOpen
  \bibfield  {author} {\bibinfo {author} {\bibfnamefont {R.}~\bibnamefont
  {Bain}}, \bibinfo {author} {\bibfnamefont {R.~P.}\ \bibnamefont {Hale}}, \
  and\ \bibinfo {author} {\bibfnamefont {S.}~\bibnamefont {Goodbred}},\
  }\href@noop {} {\bibfield  {journal} {\bibinfo  {journal} {Journal of
  Geophysical Research: Earth Surface}\ }\textbf {\bibinfo {volume} {124}},\
  \bibinfo {pages} {2141} (\bibinfo {year} {2019})}\BibitemShut {NoStop}%
\bibitem [{\citenamefont {Rodríguez-Iturbe}\ \emph
  {et~al.}(1992{\natexlab{b}})\citenamefont {Rodríguez-Iturbe}, \citenamefont
  {Rinaldo}, \citenamefont {Rigon}, \citenamefont {Bras}, \citenamefont
  {Marani},\ and\ \citenamefont {Ijjász-Vásquez}}]{rodriguez1992a}%
  \BibitemOpen
  \bibfield  {author} {\bibinfo {author} {\bibfnamefont {I.}~\bibnamefont
  {Rodríguez-Iturbe}}, \bibinfo {author} {\bibfnamefont {A.}~\bibnamefont
  {Rinaldo}}, \bibinfo {author} {\bibfnamefont {R.}~\bibnamefont {Rigon}},
  \bibinfo {author} {\bibfnamefont {R.~L.}\ \bibnamefont {Bras}}, \bibinfo
  {author} {\bibfnamefont {A.}~\bibnamefont {Marani}}, \ and\ \bibinfo {author}
  {\bibfnamefont {E.}~\bibnamefont {Ijjász-Vásquez}},\ }\href {\doibase
  10.1029/91WR03034} {\bibfield  {journal} {\bibinfo  {journal} {Water
  Resources Research}\ }\textbf {\bibinfo {volume} {28}},\ \bibinfo {pages}
  {1095} (\bibinfo {year} {1992}{\natexlab{b}})}\BibitemShut {NoStop}%
\bibitem [{\citenamefont {Banavar}\ \emph {et~al.}(2001)\citenamefont
  {Banavar}, \citenamefont {Colaiori}, \citenamefont {Flammini}, \citenamefont
  {Maritan},\ and\ \citenamefont {Rinaldo}}]{banavar2001}%
  \BibitemOpen
  \bibfield  {author} {\bibinfo {author} {\bibfnamefont {J.~R.}\ \bibnamefont
  {Banavar}}, \bibinfo {author} {\bibfnamefont {F.}~\bibnamefont {Colaiori}},
  \bibinfo {author} {\bibfnamefont {A.}~\bibnamefont {Flammini}}, \bibinfo
  {author} {\bibfnamefont {A.}~\bibnamefont {Maritan}}, \ and\ \bibinfo
  {author} {\bibfnamefont {A.}~\bibnamefont {Rinaldo}},\ }\href {\doibase
  10.1023/A:1010397325029} {\bibfield  {journal} {\bibinfo  {journal} {Journal
  of Statistical Physics}\ }\textbf {\bibinfo {volume} {104}},\ \bibinfo
  {pages} {1} (\bibinfo {year} {2001})}\BibitemShut {NoStop}%
\bibitem [{\citenamefont {Bohn}\ and\ \citenamefont
  {Magnasco}(2007)}]{bohn2007}%
  \BibitemOpen
  \bibfield  {author} {\bibinfo {author} {\bibfnamefont {S.}~\bibnamefont
  {Bohn}}\ and\ \bibinfo {author} {\bibfnamefont {M.~O.}\ \bibnamefont
  {Magnasco}},\ }\href {\doibase 10.1103/PhysRevLett.98.088702} {\bibfield
  {journal} {\bibinfo  {journal} {Phys. Rev. Lett.}\ }\textbf {\bibinfo
  {volume} {98}},\ \bibinfo {pages} {088702} (\bibinfo {year}
  {2007})}\BibitemShut {NoStop}%
\bibitem [{\citenamefont {Durand}(2007)}]{durand2007}%
  \BibitemOpen
  \bibfield  {author} {\bibinfo {author} {\bibfnamefont {M.}~\bibnamefont
  {Durand}},\ }\href {\doibase 10.1103/PhysRevLett.98.088701} {\bibfield
  {journal} {\bibinfo  {journal} {Phys. Rev. Lett.}\ }\textbf {\bibinfo
  {volume} {98}},\ \bibinfo {pages} {088701} (\bibinfo {year}
  {2007})}\BibitemShut {NoStop}%
\bibitem [{\citenamefont {Schwenk}\ and\ \citenamefont
  {Hariharan}(2021)}]{schwenk2021}%
  \BibitemOpen
  \bibfield  {author} {\bibinfo {author} {\bibfnamefont {J.}~\bibnamefont
  {Schwenk}}\ and\ \bibinfo {author} {\bibfnamefont {J.}~\bibnamefont
  {Hariharan}},\ }\href {\doibase 10.21105/joss.02952} {\bibfield  {journal}
  {\bibinfo  {journal} {Journal of Open Source Software}\ }\textbf {\bibinfo
  {volume} {6}},\ \bibinfo {pages} {2952} (\bibinfo {year} {2021})}\BibitemShut
  {NoStop}%
\bibitem [{\citenamefont {Friedrichs}\ and\ \citenamefont
  {Madsen}(1992)}]{friedrichs_nonlinear_1992}%
  \BibitemOpen
  \bibfield  {author} {\bibinfo {author} {\bibfnamefont {C.~T.}\ \bibnamefont
  {Friedrichs}}\ and\ \bibinfo {author} {\bibfnamefont {O.~S.}\ \bibnamefont
  {Madsen}},\ }\href {\doibase 10.1029/92JC00354} {\bibfield  {journal}
  {\bibinfo  {journal} {Journal of Geophysical Research: Oceans}\ }\textbf
  {\bibinfo {volume} {97}},\ \bibinfo {pages} {5637} (\bibinfo {year}
  {1992})}\BibitemShut {NoStop}%
\bibitem [{\citenamefont {Bricheno}\ \emph {et~al.}(2016)\citenamefont
  {Bricheno}, \citenamefont {Wolf},\ and\ \citenamefont
  {Islam}}]{bricheno_tidal_2016}%
  \BibitemOpen
  \bibfield  {author} {\bibinfo {author} {\bibfnamefont {L.~M.}\ \bibnamefont
  {Bricheno}}, \bibinfo {author} {\bibfnamefont {J.}~\bibnamefont {Wolf}}, \
  and\ \bibinfo {author} {\bibfnamefont {S.}~\bibnamefont {Islam}},\ }\href
  {\doibase 10.1016/j.ecss.2016.09.014} {\bibfield  {journal} {\bibinfo
  {journal} {Estuarine, Coastal and Shelf Science}\ }\textbf {\bibinfo {volume}
  {182}},\ \bibinfo {pages} {12} (\bibinfo {year} {2016})}\BibitemShut
  {NoStop}%
\bibitem [{\citenamefont {Hoitink}\ and\ \citenamefont
  {Jay}(2016)}]{hoitink_tidal_2016}%
  \BibitemOpen
  \bibfield  {author} {\bibinfo {author} {\bibfnamefont {A.~J.~F.}\
  \bibnamefont {Hoitink}}\ and\ \bibinfo {author} {\bibfnamefont {D.~A.}\
  \bibnamefont {Jay}},\ }\href {\doibase 10.1002/2015RG000507} {\bibfield
  {journal} {\bibinfo  {journal} {Reviews of Geophysics}\ }\textbf {\bibinfo
  {volume} {54}},\ \bibinfo {pages} {240} (\bibinfo {year} {2016})}\BibitemShut
  {NoStop}%
\bibitem [{\citenamefont {McLachlan}\ \emph {et~al.}(2020)\citenamefont
  {McLachlan}, \citenamefont {Ogston}, \citenamefont {Asp}, \citenamefont
  {Fricke}, \citenamefont {Nittrouer},\ and\ \citenamefont
  {Gomes}}]{mclachlan2020}%
  \BibitemOpen
  \bibfield  {author} {\bibinfo {author} {\bibfnamefont {R.}~\bibnamefont
  {McLachlan}}, \bibinfo {author} {\bibfnamefont {A.}~\bibnamefont {Ogston}},
  \bibinfo {author} {\bibfnamefont {N.}~\bibnamefont {Asp}}, \bibinfo {author}
  {\bibfnamefont {A.}~\bibnamefont {Fricke}}, \bibinfo {author} {\bibfnamefont
  {C.}~\bibnamefont {Nittrouer}}, \ and\ \bibinfo {author} {\bibfnamefont
  {V.}~\bibnamefont {Gomes}},\ }\href {\doibase
  https://doi.org/10.1016/j.ecss.2019.106524} {\bibfield  {journal} {\bibinfo
  {journal} {Estuarine, Coastal and Shelf Science}\ }\textbf {\bibinfo {volume}
  {233}},\ \bibinfo {pages} {106524} (\bibinfo {year} {2020})}\BibitemShut
  {NoStop}%
\bibitem [{\citenamefont {Seminara}\ and\ \citenamefont
  {Tubino}(2001)}]{seminara_sand_2001}%
  \BibitemOpen
  \bibfield  {author} {\bibinfo {author} {\bibfnamefont {G.}~\bibnamefont
  {Seminara}}\ and\ \bibinfo {author} {\bibfnamefont {M.}~\bibnamefont
  {Tubino}},\ }\href {\doibase 10.1017/S0022112001004748} {\bibfield  {journal}
  {\bibinfo  {journal} {Journal of Fluid Mechanics}\ }\textbf {\bibinfo
  {volume} {440}},\ \bibinfo {pages} {49} (\bibinfo {year} {2001})}\BibitemShut
  {NoStop}%
\end{thebibliography}%

\end{document}